\documentclass{aa}
\usepackage[varg]{txfonts}
\usepackage{graphicx}
\usepackage{natbib}
\usepackage{balance}
\usepackage{threeparttable}
\usepackage{multicol}
\usepackage{marginnote}
\bibpunct{(}{)}{;}{a}{}{,} 
\usepackage[normalem]{ulem}

\begin{document}

\def \jgr{J. Geophys. Res.}
\def \grl{Geophys. Res. Lett.}
\def \nat{Nature}
\def \mnras {MNRAS}
\def \aap {A{\&}A}
\def \aj {AJ}
\def \icarus {Icarus}
\def \apj {ApJ}
\def \apjl {ApJL}
\def \gca {Geochim. Cosmochim. Acta}
\def \ssr {Space Sci. Rev.}
\def \asr {Adv. Space Res.}
\def \planss {Planet. Space Sci.}
\def \dth {\frac{\partial}{\partial\theta}}
\def \dfi {\frac{\partial}{\partial\phi}}

\title{Properties of Alpha Monocerotid meteors from the observation of the 2019 outburst in the Czech Republic}
\author{Luk\'a\v{s} Shrben\'y
\and Ji\v{r}\'i Borovi\v{c}ka
\and Pavel Koten
\and Pavel Spurn\'y
\and Rostislav \v{S}tork
\and Kamil Hornoch
\and Vlastimil Voj\'a\v{c}ek
}
\institute{Astronomical Institute of the Czech Academy of Sciences, Fri\v{c}ova 298, 251\,65 Ond\v{r}ejov, Czech Republic}
\date{Received June 2021 / Accepted August 2021}
\abstract
{We observed the predicted outburst of the Alpha Monocerotid (AMO) meteor shower on 2019 November 22 with our modernized video 
and photographic cameras. Due to the short duration and moderate intensity of the outburst, atmospheric trajectories and radiants were obtained 
for only ten meteors, seven of which included velocities, magnitudes, and orbits. In addition, one incomplete video spectrum was captured. 
The radiants and orbits were found to be compatible with that of the 1995 outburst. The spectrum confirmed that AMO meteoroids are deficient 
in sodium. Unlike any other meteor shower, meteor end heights were found to be distributed along a constant level of 90\,km for all meteors 
with magnitudes between +4 and -2 and with atmospheric trajectory lengths up to 40\,km. We propose that Alpha Monocerotids were formed 
from a devolatilized and fragile cometary crust composed from relatively large fundamental grains.
}
\authorrunning{L. Shrben\'y et al.}
\titlerunning{Alpha Monocerotids 2019}
\maketitle


\section{Introduction}
The Alpha Monocerotids (AMO, IAU $\# 246$) is a meteor shower with a peak occurring around November 22 every year. The annual activity is low, but intense outbursts are occasionally produced. Five individual outbursts were observed in 1925, 1935, 1985, 1995, and 2017 \citep{jen97, jen17, rog18}. The observations in 1995 provided the first ten multi-station meteors with known radiants and heliocentric orbits giving the first evidence of the presence of a trail of a long-period comet. The peak zenith hourly rate (ZHR) of the 1995 AMO outburst was 500 \citep{jen97}. Observations of the Cameras for Allsky Meteor Surveillance (CAMS) networks in 2017 provided 17 individual orbits of AMO meteors, even if no outburst was predicted \citep{rog18}. Median orbital elements of the majority of them (10 multi-station meteors were observed in the United Arab Emirates) are in good agreement with the observations obtained in 1995 \citep{jen17}. Four television spectra of AMO meteors were obtained in 1995 by \citet{sto98}. They observed a very faint Na~I line (5892\,$\r{A}$) and concluded that the material of AMO meteoroids contains a lower amount of volatile elements, at least sodium. \citet{bor05} classified AMO meteors near the boundary between Na-poor and Na-free meteoroids according to their spectral properties. \citet{jen97} noted that Alpha Monocerotids have relatively low ends of their atmospheric trajectories compared to other cometary meteoroids with similar velocities (e.g., Perseids), and they are therefore stronger and might lack volatiles.

\citet{lyy03} found that only encounters with one-revolution trails can explain the observed outburst duration of 40~minutes. The best solution for the unknown orbital period of the parent comet was 650~years. The original calculation showed the trail will pass far from the Earth's orbit in 2019. Recently, \citet{jen19} re-evaluated this result assuming a shorter orbital period of 500~years, which moved the trail closer to the Earth. They predicted similar or even higher peak rate in comparison with the 1995 encounter. The 2019 encounter was expected at solar longitude 239.308$^\circ $; that is, on November 22, 4h~50m~UT with a full width at half maximum (FWHM) of only 0.29~hours.

Following that prediction, we decided to organize a video and digital photography campaign to observe the outburst. The campaign was successful, though the number of observed AMO meteors was not high, and we report the results here. Section \ref{so} is devoted to the description of the observations. The obtained data are described in Section \ref{sd} and analyzed from various points of view in Section \ref{sr}. The implications are discussed in Section \ref{sdi}.

\section{Observations}
\label{so}
\begin{table*}[t!]
\centering
\begin{threeparttable}
\caption{Photographic and video cameras used during AMO 2019 observation. MLM is meteor limiting magnitude, fps is frames per second. Camera codes: C is Chur\'a\v{n}ov, M Myslivna, K K\v{r}i\v{s}\v{t}anov, v is regular video camera, sp is spectral video camera, D is photographic DAFO, S is photographic SDAFO, sv is supplementary video camera, and number is focal length of supplementary photographic camera (35\,mm equivalent).}
\label{tcam}
{\tiny
\begin{tabular}{@{\extracolsep{\fill}} l|cccccc}
\hline\noalign{\smallskip}
camera  &       Resolution      &       MLM     &       fps or  &       FOV     &       Site    &       code    \\
        &       (pixels)        &       (mag)   &       sector  &       (deg)   &               &               \\
\hline\noalign{\smallskip}
\multicolumn{7}{l}{Video} \\
DMK23G445 direct        &       1280x960        &       6       &       30      &       20      &       Chur\'a\v{n}ov  &       Cv      \\
DMK23G445 spectral      &       1280x960        &       4       &       30      &       30      &       Chur\'a\v{n}ov  &       Csp     \\
DMK23G445 direct        &       1280x960        &       6       &       30      &       20      &       Myslivna        &       Mv      \\
ZWO ASI174MM    &       1936x1216       &       +1.5    &       30      &       85      &       Myslivna        &       Msv     \\
\multicolumn{7}{l}{Photographic} \\
DAFO    &       5472x3648       &       -1      &       16      &       fish-eye        &       Chur\'a\v{n}ov  &       CD      \\
SDAFO   &       5472x3648       &       1       &       -       &       diagonal fish-eye        &       Chur\'a\v{n}ov  &       CS      \\
Canon 6D        &       5472x3648       &       +0.5    &       -       &       fish-eye        &       Myslivna        &       M10     \\
Canon 450D      &       4272x2848       &       +1.5    &       -       &       70      &       Myslivna        &       M25     \\
Canon 6D II     &       6240x4160       &       +2.5    &       -       &       95      &       K\v{r}i\v{s}\v{t}anov   &       K16     \\
Canon 550D      &       5184x3456       &       +2.5    &       -       &       80      &       K\v{r}i\v{s}\v{t}anov   &       K21     \\
\hline\noalign{\smallskip}
\end{tabular}
}
\end{threeparttable}
\end{table*}
The peak timing was more favorable for Western Europe than for Central Europe; nevertheless, most of the outburst should still be visible before the beginning of dawn from Central Europe. The Moon was only 20$\%$ illuminated, so not creating much of a disturbance. Therefore, we planned to carry out the double station video observation within the Czech Republic. Due to the weather forecast, we finally selected two mountain stations in Southern Bohemia, which proved to be a lucky choice as almost the whole country was overcast that morning.

The first station was Chur\'a\v{n}ov, which is a regular station of the Czech part of the European Fireball Network (EN), equipped with Digital Autonomous Fireball Observatory (DAFO) and its spectral equivalent (SDAFO) (see \citet{spu17} and \citet{bor19} for DAFO and SDAFO descriptions, respectively). Both observatories use two Canon 6D Digital Single-Lens Reflex (DSLR) cameras to take 35 respective 30\,s long exposures of the sky. DAFO uses F3.5,~8\,mm fish-eye lens, while SDAFO uses F2.8,~15\,mm lens. DAFO is equipped with a liquid crystal shutter (LCD) for the measurement of meteor velocities. SDAFO is equipped with a holographic grating to take spectra of bright fireballs. Nevertheless, direct images (zero orders) of meteors can be used to measure meteor trajectories, though without velocity information. Thanks to its lens properties, SDAFO is more sensitive than DAFO but it does not cover the whole sky (it is a diagonal fish-eye).

For the AMO campaign, two image-intensified video cameras were additionally run at the Chur\'a\v{n}ov station, one for direct imaging and one for spectroscopy. Both instruments contain a digital DMK23G445 GigE monochromatic camera, a CCTV Basler~C125~F1.8,~8\,mm lens, and a Mullard~XX1332 image intensifier. The camera is providing a spatial resolution of 1280x960~pixels, time resolution up to 30~frames per second (fps), and a dynamic range of 8~bits \citep{kot20}. The fundamental issue was the choice of the main lens for the direct camera. We decided to use a long focal Canon~F2,~135\,mm objective lens providing high sensitivity and a field of view (FOV) of about 20~degrees in diameter. Such a relatively small FOV may result in a lower number of detections but provides precise orbits of the meteors. Because of a high predicted peak rate and recent experience with the same equipment used for the 2018 Draconid outburst observation \citep{kot20}, we estimated a recording of 10 to 20 AMO meteors, which can provide us with an insight into the comet orbit. The Jupiter F2,~85\,mm objective lens, together with a blazed spectral grating with 600 grooves per~mm, were used for the spectral video observation. The camera is sensitive to wavelengths between 3800 and 9000\,\r{A}. Again, zero-order detections could be used for trajectory and velocity measurements.

Since all other EN stations had poor weather conditions, we set up a temporary observing site at the Myslivna Mountain, about 90\,km southeast of Chur\'a\v{n}ov. That site was equipped with an identical direct imaging video camera to that of Chur\'a\v{n}ov. Spectral observations were not performed, but a supplementary video camera of type ZWO~ASI174MM with a wide FOV (about 85~degrees horizontally) was operated to record brighter meteors. For common detections with Chur\'a\v{n}ov's DAFO, another Canon~6D DSLR camera with (almost) all-sky view (10\,mm lens) was used (without LCD shutter). This camera was supplemented with a Canon\,450D camera equipped with a 25\,mm lens (35\,mm equivalent focal length) providing a FOV of 70~degrees. In addition, visual observations were performed by one of us (KH) at Myslivna.

Finally, a third backup observing site was set up at K\v{r}i\v{s}\v{t}anov, between Chur\'a\v{n}ov and Myslivna. Two DSLR cameras with 16\,mm and 21\,mm lenses (35\,mm equivalent focal length) were operated here. The parameters of all used video and photographic cameras, such as sensor pixel resolution, approximate limiting magnitude for meteors, field of view, and the frame rate (video) or shutter frequency (DAFO) are given in Table \ref{tcam}. The coordinates of the observing sites are given in Table \ref{tgps}.

The observation period was from 2:50 to 5:40~UT under clear sky throughout almost the whole period. Only high cirrus clouds shortly passed through the FOV of our video cameras. The end of the period was affected by dawn. 

The data records were inspected manually in order not to miss even very faint meteors. Meteor images and videos were measured and reduced by our standard methods \citep{spu17, kot20}.

\begin{table*}[t!]
\centering
\begin{threeparttable}
\caption{Geographical coordinates of observation sites. The comment column states the camera used.}
\label{tgps}
{\tiny
\begin{tabular}{@{\extracolsep{\fill}} l|cccc}
\hline\noalign{\smallskip}
Station &       Longitude E     &       Latitude N      &       Altitude (km)    &       comment \\
\hline\noalign{\smallskip}
\multicolumn{5}{l}{Video} \\
Chur\'a\v{n}ov  &       13.61483        &       49.06836        &       1.118   &       direct (Cv), spectral (Csp)    \\
Myslivna        &       14.68723        &       48.63044        &       1.012   &       direct (Mv), wide fov (Msv)    \\
\multicolumn{5}{l}{Photographic} \\
Chur\'a\v{n}ov  &       13.61495        &       49.06843        &       1.119   &       DAFO (CD), SDAFO (CS)        \\
Myslivna        &       14.68714        &       48.63041        &       1.012   &       all-sky (M10), wide fov (M25)   \\
K\v{r}i\v{s}\v{t}anov   &       13.99785        &       48.90794        &       0.885   &       two wide fov (K16, K21)     \\
\hline\noalign{\smallskip}
\end{tabular}
}
\end{threeparttable}
\end{table*}

\section{Data description}
\label{sd}
Three of us (JB, PS, KH) witnessed the 1995 outburst \citep{bor95a, zno95}, and our impression of the 2019 activity was definitely weaker. Nevertheless, KH, who is an experienced visual observer, counted 44~AMO meteors between 4h~26m and 5h~23m~UT, 16 of them during the ten-minute interval centered at 4h~50m~UT. The star-limiting magnitude was near 6.5 at that time.

Altogether, up to 50 different meteors were detected by at least one of the narrow field video cameras, and 18 of them were double-station cases. However, only three of them belong to the AMO shower, as measurements and calculations finally showed. Interestingly, only one of the three AMO meteors was fainter than magnitude +2.  Among the other 15 meteors, 12 were fainter than +2\,mag. We are aware that the number of meteors is not sufficient for good statistics, but it looks like there was a lack of AMO meteors fainter than +2\,mag. Our visual data contain faint AMO meteors down to magnitude +6, but in smaller proportions than for sporadic meteors.

Seven multi-station AMO meteors were recorded by photographic cameras; only one of them by DAFO, but three of them by the wide-field video (which recorded four AMO meteors in total). The velocity could therefore be computed for four photographic meteors. Photometry could be done for meteors with velocity measurements, either from DAFO or from wide-field video. The wide-field video photometry should be, nevertheless, taken as approximative.

The list of all ten multi-station AMO meteors is given in Table \ref{tatm}. The quality of the three video meteors is rather low. The meteor at 4:47:47~UT is distinct in the FOV but close to the edge of the FOV at both sites. This meteor is the best among the three. The meteor at 4:50:30~UT is very faint and hardly visible above the background, but closer to the center of the FOV. The meteor at 4:51:24~UT has the beginning out of the FOV and only the very end of the meteor is visible on the very edge of the FOV at Myslivna. This meteor would be impossible to calculate without additional data. Fortunately, the meteor was also recorded by the wide-field video camera (Msv), which has a lower level of sensitivity, in the center of its FOV. This is the reason why the beginning of this meteor, which was the brightest of the three, is at lower altitude than the other two. The photographic meteors represent a more uniform sample in this respect since all of them were recorded by SDAFO.

The spectral video camera recorded one AMO spectrum. Unfortunately, though the camera sensitivity range is from 3700 to 9000\,\r{A}, the record contains only the wavelengths from 5360 to 9000\,\r{A}, since the spectrum was out of the camera field of view for wavelengths shorter than 5360\,\r{A}. The spectrum belongs to the meteor at 5:01:51\,UT, which was observed photographically.
In the following section, we present the results of the data analysis. In particular, we concentrate on atmospheric heights, radiants, orbits, and the spectrum.

\begin{table*}[t!]
\centering
\begin{threeparttable}
\caption{Atmospheric trajectories of multi-station AMO meteors observed on 2019 November 22. $\Delta $t is an uncertainty in the beginning time, $H_{beg}$ and $H_{end}$ are observed beginning and terminal heights, $H_{begS}$ and $H_{endS}$ are beginning and terminal heights observed by a photographic spectral camera at station Chur\'a\v{n}ov (CS), L is the observed length of the atmospheric trajectory, $ZD_R$ is the zenith distance of the radiant at $H_{end}$, $M_{abs}$ is the absolute (100\,km distance) magnitude, $v_{inf}$ is the initial velocity, N is the number of stations where the meteor was observed (number in parentheses is number of stations used for the trajectory solution), and Cam refers to cameras by which the meteor was recorded (abbreviations follow the designation as in Table \ref{tcam}).}
\label{tatm}
{\tiny
\begin{tabular}{@{\extracolsep{\fill}} l|ccccccccccc}
\hline\noalign{\smallskip}
Time    &       $\Delta $t      &       $H_{beg}$       &       $H_{begS}$      &       $H_{end}$       &       $H_{endS}$      &       L       &       $ZD_R$  &       $M_{abs}$       &       $v_{inf}$       &       N       &       Cam     \\
(UT)    &       (s)     &       (km)    &       (km)    &       (km)    &       (km)    &       (km)    &       (deg)   &       (mag)   &       (km\,s$^{-1}$)  &               &               \\
\hline\noalign{\smallskip}
\multicolumn{12}{l}{Video meteors}    \\
4:47:47 &       <1      &       100.949 &       -       &       92.172  &       -       &       15.30   &       55.06   &       +2.0    &       63.7$\pm$0.3    &       2       &       Cv, Mv      \\
4:50:30 &       <1      &       100.080 &       -       &       90.660  &       -       &       16.16   &       54.42   &       +4.1    &       64.4$\pm$1.3    &       2       &       Cv, Mv      \\
4:51:24 &       <1      &       97.750$^a$      &       -       &       88.269  &       -       &       16.43   &       54.8    &       +1.2    &       62.8$\pm$0.3    &       2       &       Cv, Mv, Msv \\
\multicolumn{12}{l}{Photographic meteors}    \\
4:40:53 &       $\pm$6  &       97.898  &       97.898  &       88.985  &       90.654  &       14.68   &       52.66   &       -       &       -       &       2       &       CS, M10, M25        \\
4:44:22 &       $\pm$5  &       100.594 &       100.594 &       90.713  &       90.713  &       16.26   &       52.62   &       -       &       -       &       3       &       CS, M25, K16        \\
4:47:20 &       <1      &       95.176  &       95.126  &       90.038  &       90.741  &       8.63    &       53.83   &       -0.9    &       62.7$\pm$0.3    &       3 (2)$^b$ &       CS, Msv, K16    \\
4:58:28 &       $\pm$1  &       109.970 &       106.926 &       92.451  &       92.837  &       30.23   &       54.68   &       -2.4    &       62.93$\pm$0.12  &       3       &       CD, CS, M10, M25, K21       \\
5:01:51 &       <1      &       101.757 &       101.757 &       90.964  &       92.666  &       19.20   &       55.86   &       -1.0    &       63.2$\pm$0.4    &       2       &       CS, Csp, Msv        \\
5:07:31 &       <1      &       99.995  &       98.717  &       89.588  &       91.489  &       18.51   &       55.92   &       -0.8    &       63.0$\pm$0.3    &       2       &       CS, M10, Msv        \\
5:10:09 &       $\pm$6  &       97.134  &       95.747  &       89.732  &       89.732  &       13.25   &       56.08   &       -       &       -       &       2       &       CS, M10     \\
\hline\noalign{\smallskip}
\end{tabular}
}
{\scriptsize
\begin{tablenotes}
\item 
\textit{Notes: a) beginning of video meteor at 4:51:24~UT was not in the FOV of our regular cameras, the beginning is from the supplementary ZWO camera with lower sensitivity; b) record from K\v{r}i\v{s}\v{t}anov was not used since the meteor was recorded on the very edge of the FOV and the trajectory shows systematic shift of 70\,m in comparison to final solution.} 
\end{tablenotes}
}
\end{threeparttable}
\end{table*}

\section{Results}
\label{sr}

\subsection{Atmospheric trajectories and heights}
\label{ssa}
\begin{figure}[t]
\centering 
\includegraphics[width=0.5\textwidth]{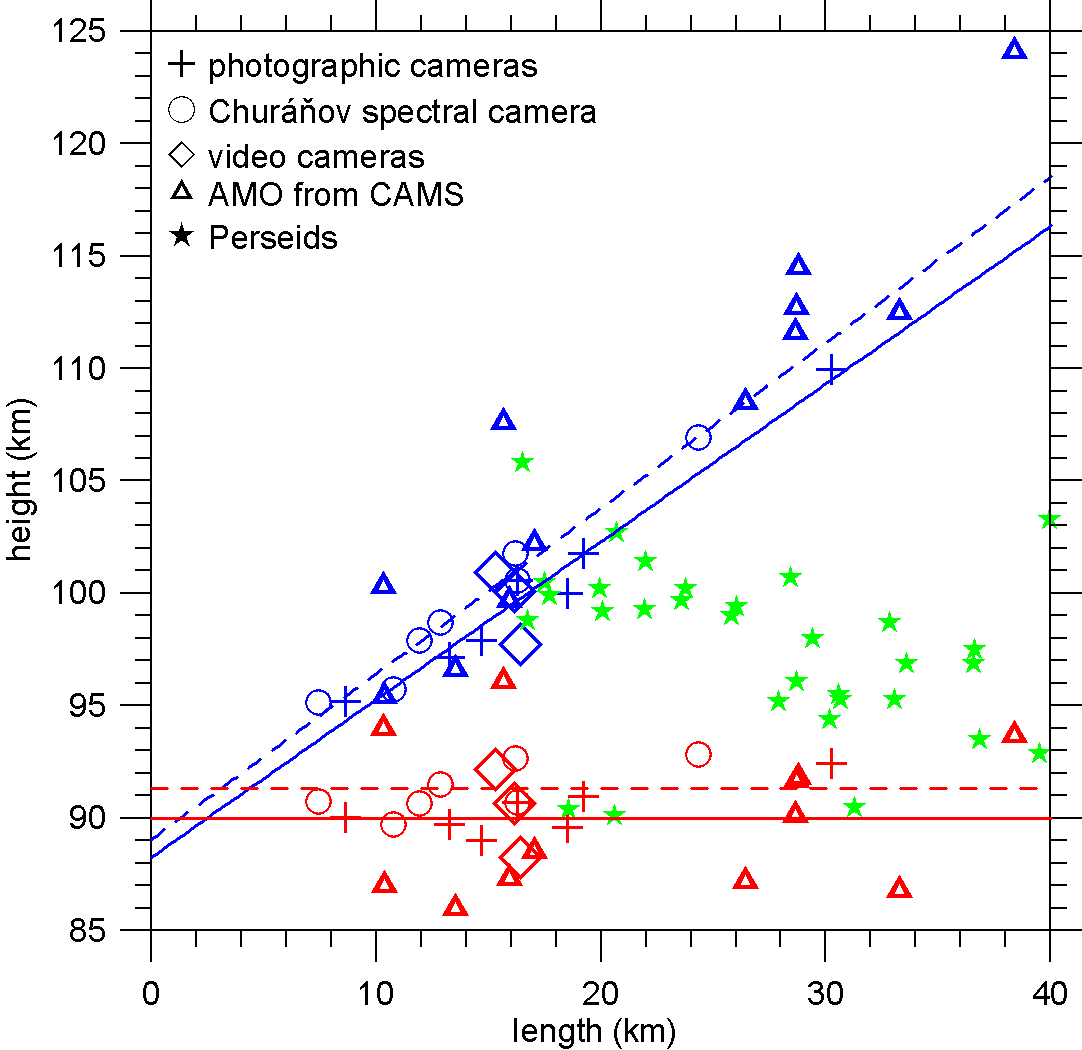}
\caption{Beginning (blue symbols) and end (red symbols) heights of the photographic meteors as a function of the length of the atmospheric trajectory. Solid lines show linear dependencies of heights $H_{beg}$ and  $H_{end}$ on observed length of atmospheric trajectory, $L$, dashed lines show linear dependencies of heights $H_{begS}$ and  $H_{endS}$ observed by spectral photographic camera at station Chur\'a\v{n}ov (CS in Table \ref{tcam}) on the length observed by the CS camera. Our three video meteors are plotted by diamonds for comparison. Values of beginning and terminal heights of AMO meteors observed by CAMS (triangles) and terminal heights of Perseid meteors (green stars) are also shown for easy comparison.}
\label{fh}
\end{figure}

Meteor beginning and end heights, trajectory lengths, and zenith distances of radiants are given in Table \ref{tatm}. Since all meteors occurred during a short time interval, their trajectory slopes are similar. Two values of beginning and end heights are listed for photographic meteors: the values from all detections, and the values based on SDAFO only. The first set depends on the type of instrument, while the second set is more homogeneous. The beginning and end heights of all AMO~2019 meteors are shown in Figure \ref{fh} as function of the length of the atmospheric trajectory. The dependencies based on SDAFO are shifted to shorter trajectories, as expected. It is reasonable to assume that higher initial mass of the meteoroids implies longer atmospheric trajectory (in case of the same velocity, slope, and material composition). We can then see the same dependency for beginning heights as that observed for other meteor showers with a cometary origin \citep[see e.g.,][]{kot04}, that is the beginning height increases with increasing meteoroid mass or is constant. The end height decreases with increasing meteoroid mass in all other showers \citep{kot04}, but it is not the case of our AMO meteors. We see slightly increasing or constant dependency. If we do not include the longest meteor 4:58:28~UT, the dependency is constant within the range of errors and the end heights are distributed along constant height 90.0$\pm$0.7\,km (if all photographic AMO meteors are used the terminal height is 90.4$\pm$1.1\,km and if also video meteors are used the height is 90.4$\pm$1.3\,km). It is improbable that the terminal height of the 4:58:28~UT meteor could be at a lower altitude, since it was recorded by the sensitive photographic camera from K\v{r}i\v{s}\v{t}anov.

It is important to note that the dependency is based on seven meteors only. So, we searched for any AMO meteor with available atmospheric trajectory for comparison. We only found a catalog of video meteors of Cameras for All-sky Meteor Surveillance (CAMS) by \citet{jen18}. The last available catalog of CAMS data is version~3, with multi-station meteors recorded between October~2010 and December~2016. We found 17 meteors designated as AMO. We then used the mean AMO orbit published by \citet{jen97} as a reference to verify the orbital similarity according to the procedure of \citet{sou63}. We removed one meteor with a value of $D_{SH} > 0.15$. Atmospheric trajectories of all our AMO meteors are shorter than 40\,km, so we only used 12 meteors from the CAMS catalog for comparison. The direct comparison would be better if only meteors with similar zenith distances of the radiant are used, but the application of this condition would lead to only one meteor for comparison. These 12 meteors are plotted in Figure \ref{fh} with triangles. The beginning heights correspond to our data, and the terminal heights are distributed along a constant height of 90$\pm$3\,km, which also corresponds to our observation. We can conclude that terminal heights of AMO meteors with atmospheric trajectories shorter than 40\,km are distributed along a constant height of 90\,km. It is important to note that there are four AMO meteors in the CAMS catalog with atmospheric trajectories longer than 40\,km. Their terminal heights decrease with increasing length of the atmospheric trajectory, as is observed for other meteor showers with cometary origin \citep{kot04}.

The terminal heights of AMO meteors shorter than 40\,km were also directly compared to the terminal heights of Perseids in Fig. \ref{fh}. Perseids have similar atmospheric velocities (60\,km\,s$^{-1}$). We selected 31 video Perseids from the data published by \citet{kot04}. Perseids with atmospheric trajectories shorter than 40\,km and zenith distances of the radiant in the 50-60 degree range (corresponding to our AMO meteors) were used. Their terminal heights decrease with increasing length of the atmospheric trajectory. Moreover, most Perseid meteors terminate at higher altitudes in comparison with AMO meteors of the same length. 

\subsection{Radiants and orbits}
\label{ssr}
\begin{figure*}
\centering 
\includegraphics[width=0.4\textwidth]{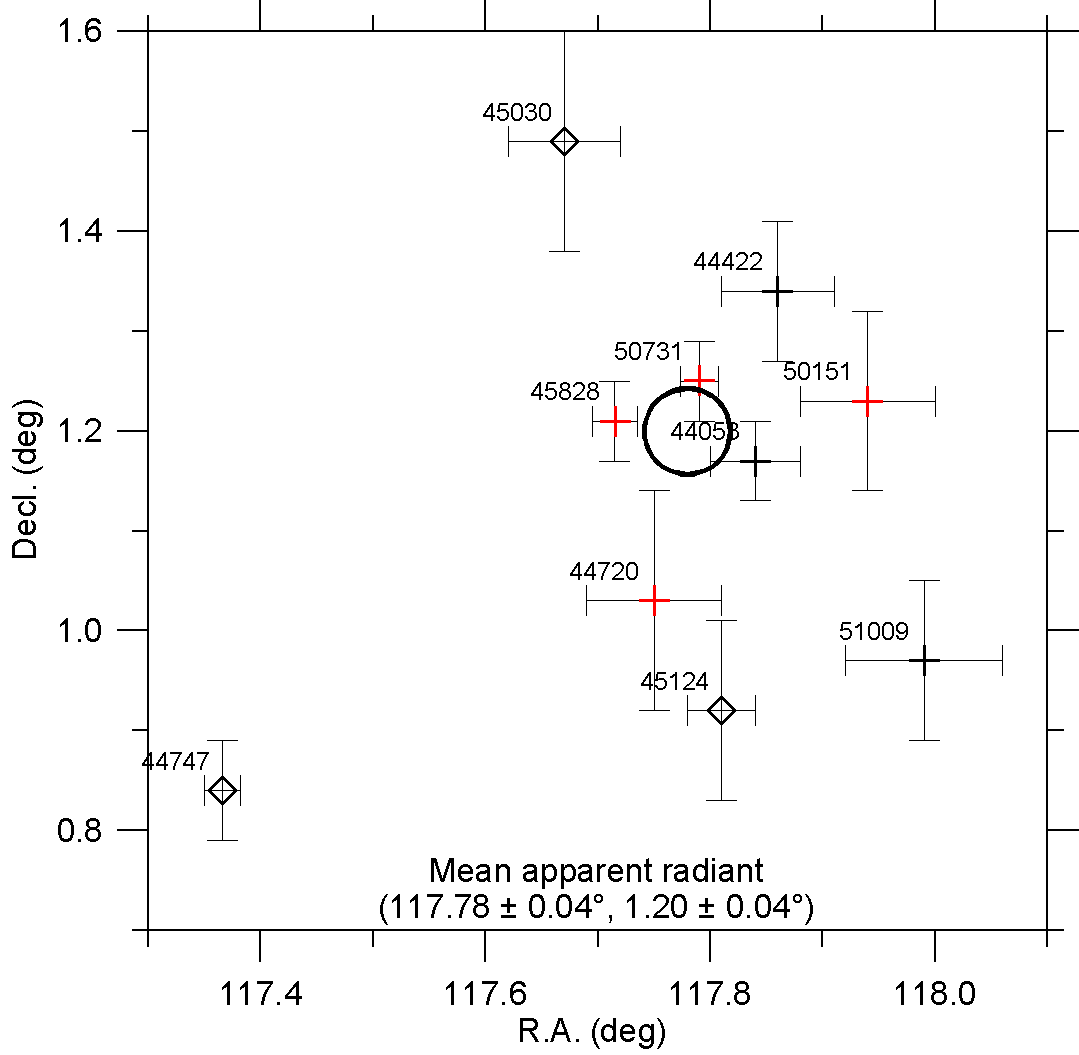}
\includegraphics[width=0.4\textwidth]{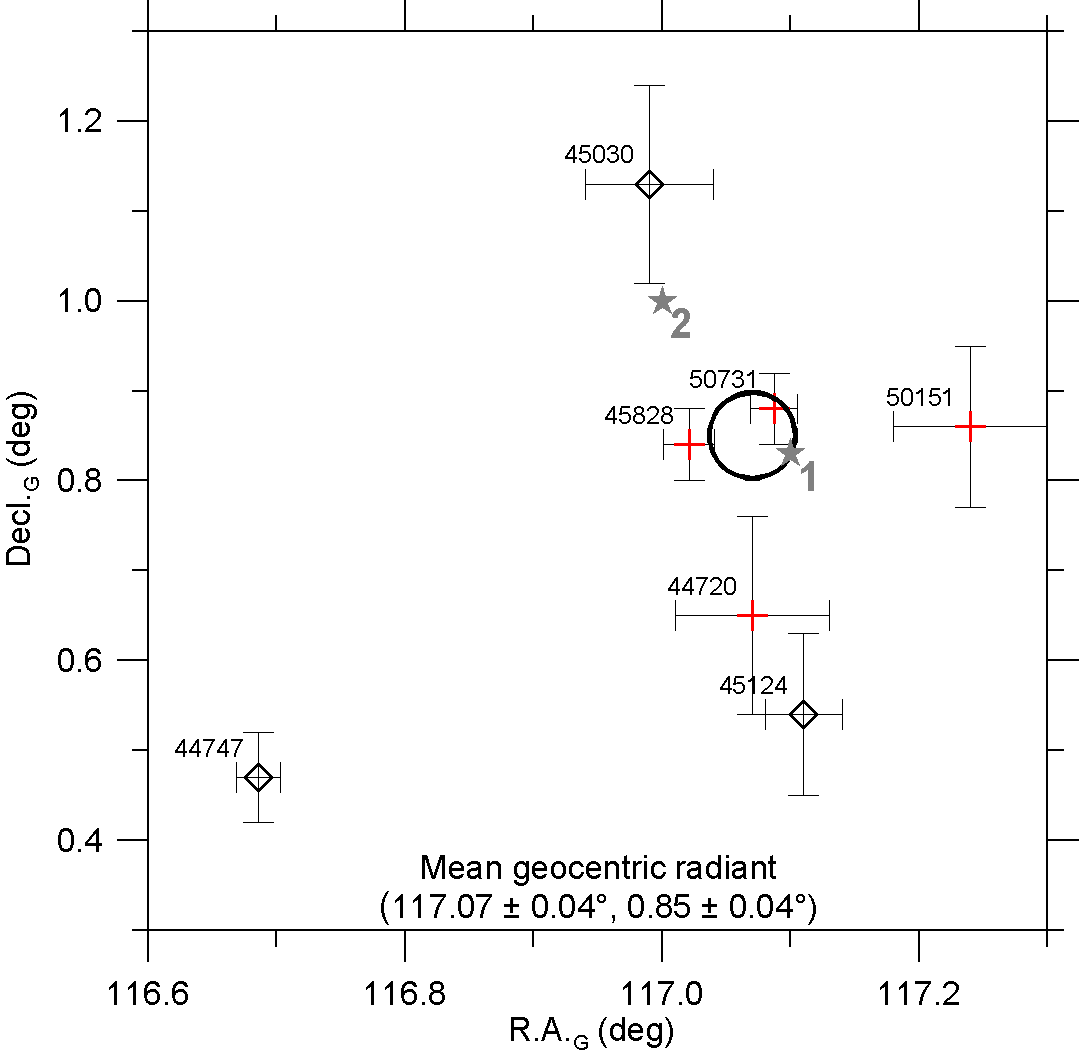}
\caption{Apparent \textit{(left panel)} and geocentric \textit{(right panel)} radiants (J2000.0) of the photographic (crosses) and video (diamonds) meteors of AMO 2019. The legend of the points is the beginning time (corresponding to data in Table \ref{tatm}). Red crosses show radiants of photographic meteors with available velocity measurements (for which the geocentric radiant and heliocentric orbit can be computed) and black circles show mean radiants (weighted average) of photographic meteors. Gray stars show geocentric radiants of sources 1 and 2 from Table \ref{tcom} for comparison. The radiant of source 3 is out of the graph borders. Sizes of black circles correspond to the uncertainty of one standard deviation.}
\label{fr}
\end{figure*}

The radiants of our AMO meteors are plotted in Figure \ref{fr} and listed in Table \ref{trad}. The area of the radiant is compact, and the mean radiant (weighted average) of the photographic meteors corresponds well to the radiant observed during the 1995 AMO outburst \citep{jen97}. The comparison of AMO radiant data published so far is presented in Table \ref{tcom}.

Heliocentric orbits could be computed for seven meteors with measured velocities. The measurement of atmospheric velocity is always challenging in the case of fast and short meteors, especially for double-station meteors. Due to a narrow field of view, we were able to determine the initial velocities of video meteors with a standard deviation bellow or slightly above 1\,km\,s$^{-1}$ (Table \ref{tatm}). The mean (weighted average) initial velocity of our three video meteors is 63.3$\pm$0.4\,km\,s$^{-1}$.

Atmospheric velocities of photographic meteors are based on two different instruments. The meteor at 4:58:28~UT was recorded by DAFO, and the other three meteors were recorded by the wide-field video camera. Short atmospheric trajectories lead to the initial velocities with a standard deviation of few tenths of km\,s$^{-1}$ (Table \ref{tatm}). The mean (weighted average) initial velocity of our four photographic meteors is 62.99$\pm$0.06\,km\,s$^{-1}$. The weighted average initial velocity of all seven AMO meteors is 63.00$\pm$0.15\,km\,s$^{-1}$.

Orbital elements of all AMO 2019 meteors and their weighted averages are presented in Table \ref{trad}. It was not possible to determine a reliable semi-major axis on the basis of such small number of short and fast meteors. Even if the radiant is well determined, the uncertainty in initial velocity (velocities 62.7-64.4\,km\,s$^{-1}$) results in orbits of either short-period or long-period comets or even hyperbolic orbits. From the nature of AMO outbursts, it is known that only encounters with a one-revolution trail of an unknown long-period comet can explain the shower \citep{lyy03}, so only elliptical orbits are possible. The weighted average value of the semi-major axis is 21$\pm$6\,AU, which gives an orbital period of around 100~years. However, longer periods would be consistent with our data as well, so this can be taken as a lower limit. \citet{jen06} also used the available AMO observations to determine the lower limit of the semi-major axis. He concluded that a~>~28\,AU (period~>~150~years).

\begin{table*}[t!]
\centering
\begin{threeparttable}
\caption{Radiants and orbital elements of multi-station AMO meteors observed on 2019 November 22.}
\label{trad}
{\tiny
\begin{tabular}{@{\extracolsep{\fill}} l|cccccccccc}
\hline\noalign{\smallskip}
Time    &       R.A.    &       Decl.   &       R.A.$_G$        &       Decl.$_G$       &       v$_G$   &       e       &       q       &       $\omega$        &       i       &       T$_J$   \\
(UT)    &       (deg)   &       (deg)   &       (deg)   &       (deg)   &       (kms$^{-1}$)    &               &       (AU)    &       (deg)   &       (deg)   &               \\
\hline\noalign{\smallskip}
\multicolumn{11}{l}{Video meteors} \\
4:47:47 &       117.366 &       0.84    &       116.686 &       0.47    &       62.9    &       1.016   &       0.491   &       89.8    &       133.21  &       -0.77   \\
        &       $\pm$0.016      &       $\pm$0.05       &       $\pm$0.017      &       $\pm$0.05       &       $\pm$0.3        &       $\pm$0.015      &       $\pm$0.006      &       $\pm$1.1        &       $\pm$0.26       &               \\
4:50:30 &       117.67  &       1.49    &       116.99  &       1.13    &       63.6    &       1.04    &       0.495   &       89      &       135.1   &       -1.0    \\
        &       $\pm$0.05       &       $\pm$0.11       &       $\pm$0.05       &       $\pm$0.11       &       $\pm$1.3        &       $\pm$0.06       &       $\pm$0.021      &       $\pm$4  &       $\pm$0.9        &               \\
4:51:24 &       117.81  &       0.92    &       117.11  &       0.54    &       62.0    &       0.969   &       0.481   &       92.4    &       133.17  &       -0.25   \\
        &       $\pm$0.03       &       $\pm$0.09       &       $\pm$0.03       &       $\pm$0.09       &       $\pm$0.3        &       $\pm$0.012      &       $\pm$0.005      &       $\pm$1.0        &       $\pm$0.27       &               \\
WA      &       117.48  &       0.94    &       116.81  &       0.57    &       62.5    &       0.989   &       0.485   &       91.1    &       133.3   &       -0.5    \\
        &       $\pm$0.16       &       $\pm$0.18       &       $\pm$0.16       &       $\pm$0.19       &       $\pm$0.4        &       $\pm$0.022      &       $\pm$0.004      &       $\pm$1.1        &       $\pm$0.5        &       $\pm$0.7        \\
\hline\noalign{\smallskip}      
\multicolumn{11}{l}{Photographic meteors.}\\
4:40:53 &       117.84  &       1.17    &               &               &               &               &               &               &               &               \\
        &       $\pm$0.04       &       $\pm$0.04       &               &               &               &               &               &               &               &               \\
4:44:22 &       117.86  &       1.34    &               &               &               &               &               &               &               &               \\
        &       $\pm$0.05       &       $\pm$0.07       &               &               &               &               &               &               &               &               \\
4:47:20 &       117.75  &       1.03    &       117.07  &       0.65    &       61.9    &       0.967   &       0.477   &       92.9    &       133.3   &       -0.22   \\
        &       $\pm$0.06       &       $\pm$0.11       &       $\pm$0.06       &       $\pm$0.011      &       $\pm$0.3        &       $\pm$0.011      &       $\pm$0.005      &       $\pm$0.9        &       $\pm$0.3        &               \\
4:58:28 &       117.715 &       1.21    &       117.021 &       0.84    &       62.11   &       0.975   &       0.4767  &       92.7    &       133.65  &       -0.32   \\
        &       $\pm$0.020      &       $\pm$0.04       &       $\pm$0.020      &       $\pm$0.04       &       $\pm$0.12       &       $\pm$0.005      &       $\pm$0.0022     &       $\pm$0.4        &       $\pm$0.11       &               \\
5:01:51 &       117.94  &       1.23    &       117.24  &       0.86    &       62.4    &       0.983   &       0.484   &       91.6    &       134.1   &       -0.4    \\
        &       $\pm$0.06       &       $\pm$0.09       &       $\pm$0.06       &       $\pm$0.09       &       $\pm$0.4        &       $\pm$0.018      &       $\pm$0.008      &       $\pm$1.4        &       $\pm$0.4        &               \\
5:07:31 &       117.790 &       1.25    &       117.087 &       0.88    &       62.2    &       0.977   &       0.478   &       92.5    &       133.84  &       -0.35   \\
        &       $\pm$0.017      &       $\pm$0.04       &       $\pm$0.018      &       $\pm$0.04       &       $\pm$0.3        &       $\pm$0.012      &       $\pm$0.005      &       $\pm$1.0        &       $\pm$0.23       &               \\
5:10:09 &       117.99  &       0.97    &               &               &               &               &               &               &               &               \\
        &       $\pm$0.07       &       $\pm$0.08       &               &               &               &               &               &               &               &               \\
WA      &       117.78  &       1.20    &       117.07  &       0.85    &       62.11   &       0.9745  &       0.4773  &       92.64   &       133.67  &       -0.31   \\
        &       $\pm$0.04       &       $\pm$0.04       &       $\pm$0.04       &       $\pm$0.04       &       $\pm$0.07       &       $\pm$0.0021     &       $\pm$0.0013     &       $\pm$0.20       &       $\pm$0.10       &       $\pm$0.11       \\
\hline\noalign{\smallskip}      
\multicolumn{11}{l}{All meteors}\\
WA      &       117.66  &       1.15    &       116.95  &       0.77    &       62.18   &       0.977   &       0.4788  &       92.4    &       133.59  &       -0.34   \\
        &       $\pm$0.08       &       $\pm$0.05       &       $\pm$0.08       &       $\pm$0.07       &       $\pm$0.15       &       $\pm$0.007      &       $\pm$0.0025     &       $\pm$0.5        &       $\pm$0.13       &       $\pm$0.19       \\
\hline\noalign{\smallskip}
\end{tabular}
}
{\scriptsize
\begin{tablenotes}
\item 
\textit{Note: WA stands for weighted average. The uncertainties of one standard deviation are given in rows below the values.} 
\end{tablenotes}
}
\end{threeparttable}
\end{table*}

\begin{table*}[t!]
\centering
\begin{threeparttable}
\caption{Mean values of geocentric radiants and orbital elements of AMO meteors published so far.}
\label{tcom}
{\tiny
\begin{tabular}{@{\extracolsep{\fill}} l|cccccccccc}
\hline\noalign{\smallskip}
Source  &       R.A.$_G$        &       Decl.$_G$       &       v$_G$   &       a       &       e       &       q       &       $\omega$        &       $\Omega$        &       i       &       T$_J$   \\
        &       (deg)   &       (deg)   &       (km\,s$^{-1}$)  &       (AU)    &               &       (AU)    &       (deg)   &       (deg)   &       (deg)   &               \\
\hline\noalign{\smallskip}      
1       &       117.10  &       0.83    &       63.0    &       500     &       0.999   &       0.488   &       90.7    &       59.3    &       134.1   &       -0.60   \\
2       &       117.0   &       1.0     &       62.6    &       43      &       0.989   &       0.474   &       92      &       59.527  &       134.0   &       -0.49   \\
3       &       117.53  &       1.18    &       62.6    &               &               &       0.485   &       91.25   &       59.425  &       138.18  &               \\
4       &       117.07  &       0.85    &       62.11   &       18.7    &       0.9745  &       0.4773  &       92.64   &       59.310  &       133.67  &       -0.31   \\
5       &       116.81  &       0.57    &       62.5    &       44      &       0.989   &       0.485   &       91.1    &       59.307  &       133.3   &       -0.47   \\
6       &       116.95  &       0.77    &       62.18   &       21      &       0.977   &       0.4788  &       92.4    &       59.307  &       133.59  &       -0.34   \\
\hline\noalign{\smallskip}
\end{tabular}
}
{\scriptsize
\begin{tablenotes}
\item 
\textit{Note: Reference 1 is \citet{jen97}; 2 is \citet{jen17}; 3 is the annual component from \citet{rog18}; 4 is from photographic data of this work; 5 is from video data of this work; 6 is from both video and photographic data of this work.} 
\end{tablenotes}
}
\end{threeparttable}
\end{table*}

\subsection{The spectrum}
\label{sss}
\begin{figure}
\centering 
\includegraphics[width=0.5\textwidth]{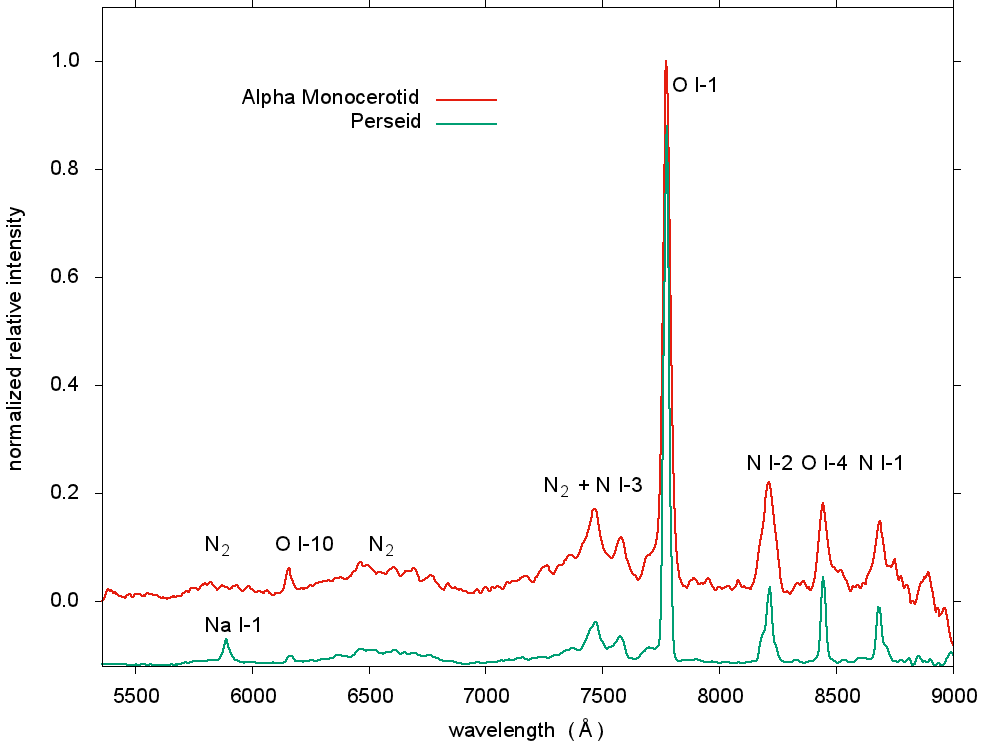}
\caption{Comparison between normalized calibrated spectra of Perseid and Alpha Monocerotid (5:01:51~UT) meteors. The spectrum of Perseid is shifted by 0.1 in intensity for clarity. No sodium Na~I-1 and only weak molecular radiation of nitrogen N$_2$ can be seen in the region between 5600 and 6100\,\r{A} for the Alpha Monocerotid spectrum. The sodium is visible in the Perseid spectrum.}
\label{fspe}
\end{figure}

The single recorded spectrum is shown in Figure \ref{fspe}, where it is also compared with the spectrum of one Perseid recorded with the same equipment. In the covered spectral range, the line of sodium Na~I at 5892\,\r{A} is usually dominant along with the atmospheric lines of O~I and N~I and N$_2$ in the region between 7000 and 8500\,\r{A}, especially for fast meteors \citep{voj15}. Due to similar velocities, the relative intensities of atmospheric lines are similar for both spectra. The sodium, as the only significant meteoritic line visible in this spectral range, is present in the Perseid spectrum, but it is missing from the AMO spectrum. This is in an agreement with the work of \citet{sto98}, as they observed a low amount of Na in the spectra of members of the AMO outburst on 1995 November 22. \citet{bor05} concluded that AMO meteors fall near the boundary between Na-poor and Na-free meteoroids according to their spectral properties (Mg-Na-Fe ratios). 

\section{Discussion}
\label{sdi}
We can confirm that the predicted outburst of the Alpha Monocerotid meteor shower occurred on 2019 November 22. The number of recorded meteors is not sufficient to compute an activity profile or meteoroid flux. The data of sensitive video cameras suggest a deficit of meteors fainter than magnitude +2, but this result has low statistical significance.  It would suggest a small population index, but $r = 3.00 \pm 0.18$ was reported by \citet{mis20} from visual data.  On the other hand, the radio data from Belarus reported by \citet{rog20} showed only a small enhancement of echoes during the outburst, indicating that the outburst was indeed inconspicuous at small masses.

The computed radiants and heliocentric orbits match those reported for the 1995 and 2017 outbursts. The radiant of the annual component reported by \citet{rog18} lies at a somewhat larger right ascension and declination. The radiant in 2019 was quite compact, with a diameter of less than one degree. The semi-major axis and period of the shower and the parent comet remains unknown. Our nominal mean solution of the period is only 100~years, but the uncertainty is large, and much longer periods are possible as well.

The most important results come from spectroscopy and peculiar tendency of terminal heights. Our earlier observation that AMO meteoroids are deficient in sodium \citep{sto98, bor05} was confirmed. \citet{bor05} identified three populations of Na-free meteoroids. The first population is of iron meteoroids, and the second one contains Sun-approaching meteoroids with small perihelia. Alpha Monocerotids do not belong to either of these. They belong to the third population, which feature meteoroids in Halley-type orbits. Na-free meteoroids in Halley-type orbits are mostly sporadic. Alpha Monocerotids is the only known meteor shower of this type. \citet{bor05} proposed that these meteoroids originate from primordial cometary crusts, which were exposed to cosmic rays for a long period of time, and volatile elements including sodium were lost.

A new finding is that small and medium-sized AMO meteoroids finish their ablation at the same height in the atmosphere (near 90\,km). This is unusual because in other meteor showers terminal height decreases with increasing meteoroid mass and trajectory length. The beginning height usually increases with mass, except for Geminids, where it is nearly constant \citep{kot04}. Geminids are also partly deficient in sodium because of their proximity to the Sun. But Alpha Monocerotids behave differently, with the beginning height increasing and terminal height being constant, demonstrating that their physical structure is different from any major meteoroid stream. 

The meteor at 4:58:28~UT is our brightest and best observed AMO 2019 meteor. It was recorded from three sites and also by DAFO at Chur\'a\v{n}ov station, so it was possible to process this meteor using our standard procedures \citep{bor90, bor95b, cep87}. We determined the value of a PE coefficient that describes the fireball ablation ability \citep{cep76}. The value of PE$ = -5.31$ lies on the border between types~II and IIIA, which corresponds to meteoroids composed of carbonaceous chondrites or regular cometary material \citep{cep88}. Moreover, it is similar to the mean PE value of Perseids and September epsilon Perseids \citep{shr19}. This meteor behaves like ordinary cometary meteoroids according to the PE criterion. However, this may just be a coincidence since its end height is not too different from Perseids of the same length (Fig. \ref{fh}). Moreover, the PE criterion was originally developed for bright fireballs and may not be a fully appropriate measure for smaller meteoroids. The photometric mass of the 4:58:28~UT meteor was only $10^{-4}$\,kg.

\begin{figure}
\centering 
\includegraphics[width=0.5\textwidth]{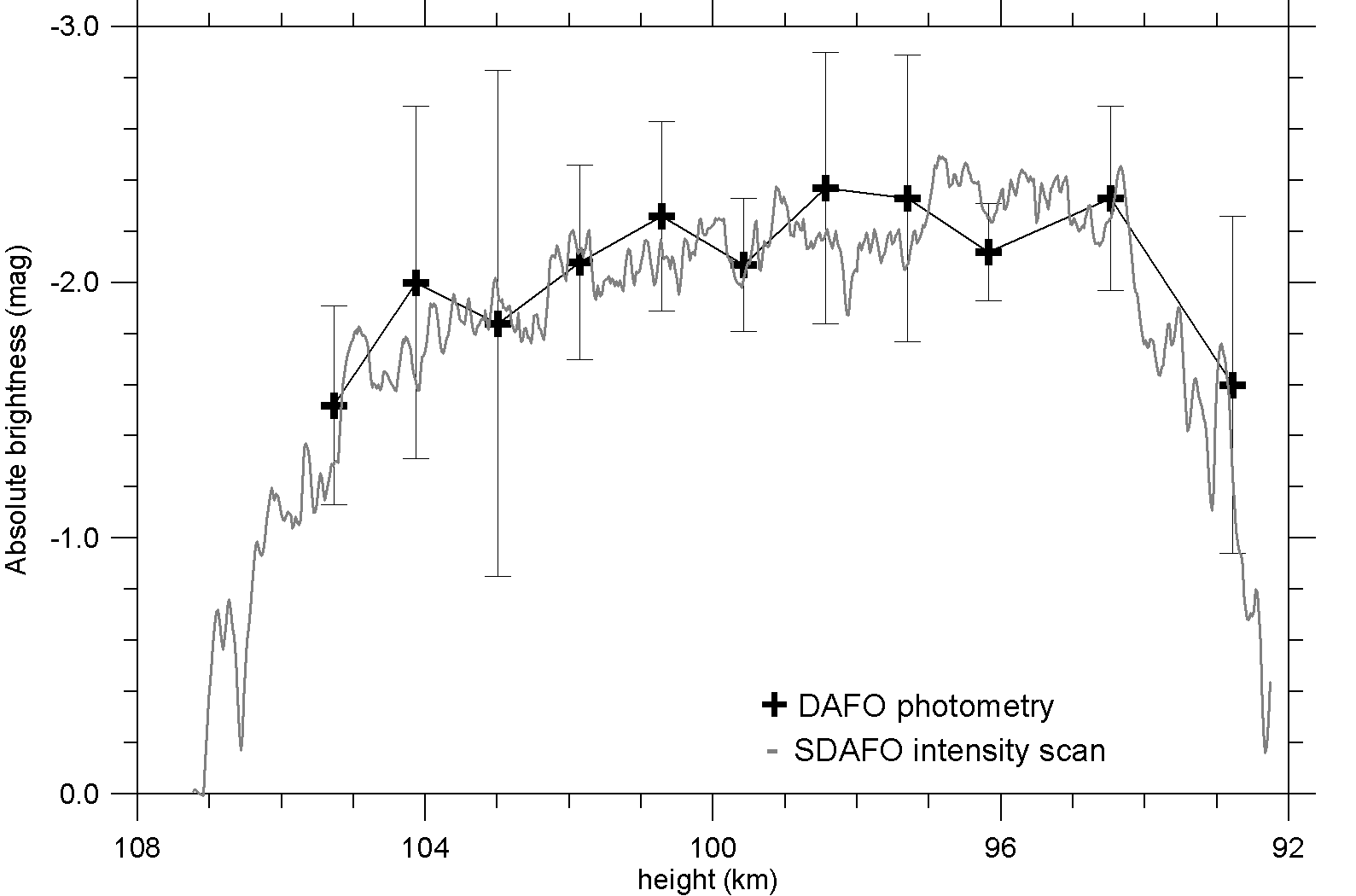}
\caption{Light curve of the brightest AMO meteor (4:58:28~UT) computed from DAFO photometry. The intensity scan on the SDAFO image was scaled to match the light curve. Small-scale variations are due to noise.}
\label{flc}
\end{figure}

The light curve of meteor at 4:58:28~UT is shown in Figure \ref{flc}. Black crosses correspond to absolute (100\,km distance) brightness from DAFO at Chur\'a\v{n}ov, and the gray curve corresponds to an intensity scan of the meteor on a spectral image from SDAFO at Chur\'a\v{n}ov. The meteor was not bright enough to be recorded by the DAFO radiometer. The light curve is without any significant flare, which corresponds to a material that did not break down suddenly into a large number of small grains in the later part of the trajectory (as bright Perseids often do). Nevertheless, the onset of the brightness is quite steep. The middle of the light curve is rather flat, with only a small increase in brightness. The fading at the end is steep again. Such a light curve does not correspond to a single ablating body, which would exhibit a steady brightness increase in the first three quarters of the trajectory and then somewhat steeper fading. The observed light curve could be explained by a disruption of the meteoroid into a large number of grains of similar sizes at the beginning, which then ablate independently and are all exhausted at similar heights, causing the sudden end of the meteor. 

We therefore propose that AMO meteoroids are devolatilized but also very crumbly. Meteoroids of all sizes disrupt into grains at the beginning of the atmospheric entry. The grain size is a typical material property, so different meteoroids disrupt into similar grains. The end height depends on grain properties and does therefore not depend on the initial meteoroid size, at least up to a critical size. Larger meteoroids just produce more grains, which leads to greater brightness and earlier detection of the meteor. Meteoroids larger than the critical size may contain some larger grains or take longer to disrupt completely, so their end heights are progressively lower.

The comparison with small Perseids shows that Alpha Monocerotid grains must be more refractory than the Perseid material. The lack of volatiles causes the ablation temperature to be higher, and the beginning heights of Alpha Monocerotids are therefore lower. Perseids surely also fragment into grains, though possibly in a different way. Yet, the AMO grains must be either larger, denser, or more resistant, and they penetrate deeper. In this sense, the hypothesis of \citet{jen97} that Alpha Monocerotids are stronger compared to other cometary meteoroids can be confirmed.

\begin{figure}
\centering 
\includegraphics[width=0.5\textwidth]{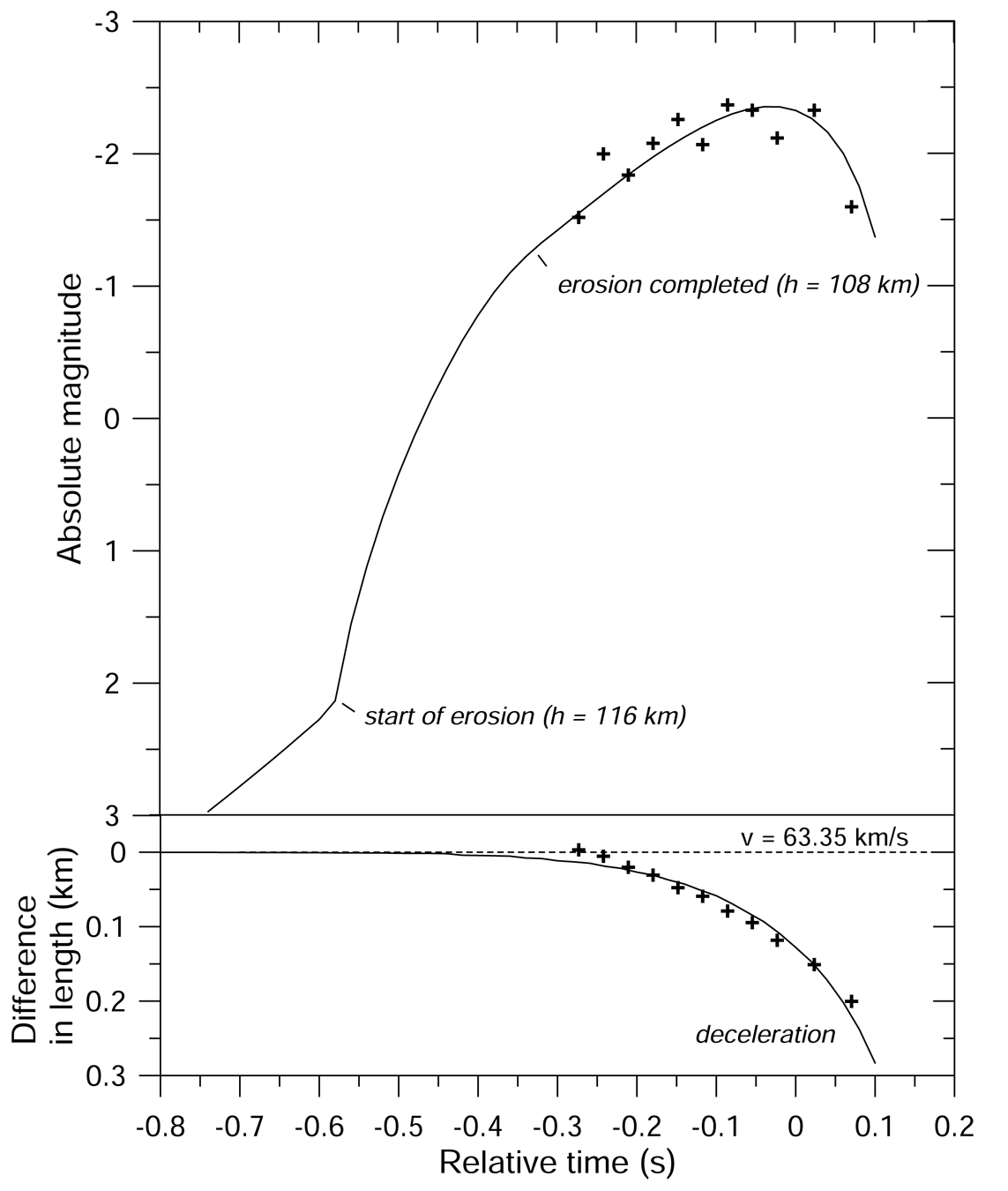}
\caption{Comparison of observed and modeled light curves \textit{(upper panel)} and deceleration \textit{(lower panel)} of the 4:58:28\,UT meteor. Observed data are given as crosses, and the model is shown with smooth solid lines. Deceleration is expressed as the difference of the length along the trajectory (of the brightest part of meteor streak) at the given time from the length expected for constant meteor velocity of 63.35\,km\,s$^{-1}$. CIRA72 model atmosphere was used to compute the drag.}
\label{fmod}
\end{figure}

To get better insight into possible properties of AMO grains, we tried to model the light curve of the 4:58:28~UT meteor with the erosion model of \citet{bor07}, originally used for faint Draconid meteors. A model assuming disruption into grains of masses of about $10^{-8}$\,kg near the beginning of the meteor was able to reproduce the light curve, but it produced a high deceleration of the meteor. The dynamics could be brought into rough agreement with the data if the grain release (erosion) started well before the observed beginning of the meteor (at heights around 115\,km or more) and was finished near the meteor's beginning (Figure \ref{fmod}). In that model, the grains were already partly decelerated at the observed meteor's beginning, so the actual entry speed was not the measured 62.93$\pm$0.12\,km\,s$^{-1}$ given in Table \ref{tatm}, but 63.35$\pm$0.30\,km\,s$^{-1}$. The corresponding eccentricity would be 0.992$\pm$0.013, corresponding to a semi-major axis of 65~AU and an orbital period of 500~years, with large uncertainties. 

The grain masses were about $2\times 10^{-7}$\,kg (uncertainty of about~50\,$\%$), corresponding to a diameter of about 0.5\,mm for an assumed density of 3000\,kg\,m$^{-3}$. Such grain sizes are larger than for the vast majority of cometary meteoroids modeled with the erosion model \citep{bor07, bor14, cam13, voj19}. We note that the lack of small grains was found as a common property for Na-poor and Na-free meteoroids \citep{voj19}. The ablation coefficient was found to be $\approx 0.025$\, s$^{2}$\,km$^{-2}$ (uncertainty of about~25\,$\%$), which is comparable with Draconids. The 4:58:28~UT meteor disrupted into $\approx 500$ grains (assuming the luminous efficiency of \citet{pec83}). A single grain would produce meteor with maximum absolute magnitude of about +4.

The fact that the parent comet of Alpha Monocerotids has not been yet discovered may be simply because of its long period. Nevertheless, it is also possible that the comet is only weakly active and is still largely covered by its primordial crust.

\section {Conclusions}
\label{sc}
We presented atmospheric trajectories and radiants of ten Alpha Monocerotid meteors observed by photographic and video techniques during the outburst on 2019 November 22. Velocities, magnitudes, and orbits were obtained for seven~meteors, and one incomplete video spectrum was captured. The AMO activity was observed between 4h~26m and 5h~23m~UT, with maximum rates from visual counts in the ten-minute interval centered at 4h~50m~UT. The geocentric radiant was compact, with a diameter of less than one degree, and was centered at R.A.$_G = 116.95\pm0.08^{\circ }$, Decl.$_G = 0.77\pm0.07^{\circ }$ (one sigma errors). The geocentric velocity was found to be $62.18\pm0.15$\,km\,s$^{-1}$. The orbital period is 100~years, or possibly longer if the measured velocities were affected by deceleration, as it was found from detailed modeling of the brightest recorded AMO.

The most interesting aspects are the physical properties of Alpha Monocerotid meteoroids, which were found to be distinct from any other meteoroid stream analyzed so far. The observed meteor end heights were found to be distributed along a constant level of 90\,km for all meteors with magnitudes between +4 and -2 and with atmospheric trajectory lengths up to 40\,km. We also confirmed that AMO meteoroids are deficient in sodium. We propose that Alpha Monocerotids were formed from a devolatilized cometary crust that is structurally fragile and disintegrates into fundamental grains of a characteristic size of about 0.5\,mm at the beginning of atmospheric entry. These grains are larger than grains of typical cometary meteoroids derived from active comets such as Perseids or Draconids.

\begin{acknowledgements}
We thank the staff of the Chur\'a\v{n}ov meteorological station operated by the Czech Hydrometeorological Institute for their assistance with the observing campaign. This work was supported by the institutional project RVO: 67985815 and grant no. 19-26232X from the Czech Science Foundation.
\end{acknowledgements}
\bibliographystyle{aa} 
\bibliography{mybibfile.bib} 

\begin{thebibliography}{30}
\expandafter\ifx\csname natexlab\endcsname\relax\def\natexlab#1{#1}\fi

\bibitem[{{Borovi{\v{c}}ka}(1990)}]{bor90}
{Borovi{\v{c}}ka}, J. 1990, Bulletin of the Astronomical Institutes of
  Czechoslovakia, 41, 391

\bibitem[{{Borovi{\v{c}}ka} {et~al.}(2014){Borovi{\v{c}}ka}, {Koten},
  {Shrben{\'y}}, {{\v{S}}tork}, \& {Hornoch}}]{bor14}
{Borovi{\v{c}}ka}, J., {Koten}, P., {Shrben{\'y}}, L., {{\v{S}}tork}, R., \&
  {Hornoch}, K. 2014, Earth Moon and Planets, 113, 15

\bibitem[{{Borovi{\v{c}}ka} {et~al.}(2005){Borovi{\v{c}}ka}, {Koten},
  {Spurn{\'y}}, {Bo{\v{c}}ek}, \& {{\v{S}}tork}}]{bor05}
{Borovi{\v{c}}ka}, J., {Koten}, P., {Spurn{\'y}}, P., {Bo{\v{c}}ek}, J., \&
  {{\v{S}}tork}, R. 2005, \icarus, 174, 15

\bibitem[{{Borovi{\v{c}}ka} \& {Spurn{\'y}}(1995)}]{bor95a}
{Borovi{\v{c}}ka}, J. \& {Spurn{\'y}}, P. 1995, WGN, Journal of the
  International Meteor Organization, 23, 203

\bibitem[{{Borovi{\v{c}}ka} {et~al.}(1995){Borovi{\v{c}}ka}, {Spurn{\'y}}, \&
  {Kecl{\'i}kov{\'a}}}]{bor95b}
{Borovi{\v{c}}ka}, J., {Spurn{\'y}}, P., \& {Kecl{\'i}kov{\'a}}, J. 1995,
  \aaps, 112, 173

\bibitem[{{Borovi{\v{c}}ka} {et~al.}(2007){Borovi{\v{c}}ka}, {Spurn{\'y}}, \&
  {Koten}}]{bor07}
{Borovi{\v{c}}ka}, J., {Spurn{\'y}}, P., \& {Koten}, P. 2007, \aap, 473, 661

\bibitem[{{Borovi{\v{c}}ka} {et~al.}(2019){Borovi{\v{c}}ka}, {Spurn{\'y}}, \&
  {Shrben{\'y}}}]{bor19}
{Borovi{\v{c}}ka}, J., {Spurn{\'y}}, P., \& {Shrben{\'y}}, L. 2019, in
  International Meteor Conference 2018, ed. R.~{Rudawska}, J.~{Rendtel},
  C.~{Powell}, R.~{Lunsford}, C.~{Verbeeck}, \& A.~{Kn{\"{o}}fel}, 28--32

\bibitem[{{Campbell-Brown} {et~al.}(2013){Campbell-Brown}, {Borovi{\v{c}}ka},
  {Brown}, \& {Stokan}}]{cam13}
{Campbell-Brown}, M.~D., {Borovi{\v{c}}ka}, J., {Brown}, P.~G., \& {Stokan}, E.
  2013, \aap, 557, A41

\bibitem[{{Ceplecha}(1987)}]{cep87}
{Ceplecha}, Z. 1987, Bulletin of the Astronomical Institutes of Czechoslovakia,
  38, 222

\bibitem[{{Ceplecha}(1988)}]{cep88}
{Ceplecha}, Z. 1988, Bulletin of the Astronomical Institutes of Czechoslovakia,
  39, 221

\bibitem[{{Ceplecha} \& {McCrosky}(1976)}]{cep76}
{Ceplecha}, Z. \& {McCrosky}, R.~E. 1976, \jgr, 81, 6257

\bibitem[{{Jenniskens}(2006)}]{jen06}
{Jenniskens}, P. 2006, {Meteor Showers and their Parent Comets} ({Cambridge
  University Press})

\bibitem[{{Jenniskens} {et~al.}(2018){Jenniskens}, {Baggaley}, {Crumpton},
  {Aldous}, {Pokorny}, {Janches}, {Gural}, {Samuels}, {Albers}, {Howell},
  {Johannink}, {Breukers}, {Odeh}, {Moskovitz}, {Collison}, \& {Ganju}}]{jen18}
{Jenniskens}, P., {Baggaley}, J., {Crumpton}, I., {et~al.} 2018, \planss, 154,
  21

\bibitem[{{Jenniskens} {et~al.}(1997){Jenniskens}, {Betlem}, {de Lignie}, \&
  {Langbroek}}]{jen97}
{Jenniskens}, P., {Betlem}, H., {de Lignie}, M., \& {Langbroek}, M. 1997, \apj,
  479, 441

\bibitem[{{Jenniskens} \& {Lyytinen}(2019)}]{jen19}
{Jenniskens}, P. \& {Lyytinen}, E. 2019, Central Bureau Electronic Telegrams,
  4692, 1

\bibitem[{{Jenniskens} \& {Odeh}(2017)}]{jen17}
{Jenniskens}, P. \& {Odeh}, M. 2017, Central Bureau Electronic Telegrams, 4457,
  1

\bibitem[{{Koten} {et~al.}(2004){Koten}, {Borovi{\v{c}}ka}, {Spurn{\'y}},
  {Betlem}, \& {Evans}}]{kot04}
{Koten}, P., {Borovi{\v{c}}ka}, J., {Spurn{\'y}}, P., {Betlem}, H., \& {Evans},
  S. 2004, \aap, 428, 683

\bibitem[{{Koten} {et~al.}(2020){Koten}, {Borovi{\v{c}}ka},
  {Voj{\'a}{\v{c}}ek}, {Spurn{\'y}}, {{\v{S}}tork}, {Shrben{\'y}}, {Janout},
  {Fliegel}, {P{\'a}ta}, \& {V{\'\i}tek}}]{kot20}
{Koten}, P., {Borovi{\v{c}}ka}, J., {Voj{\'a}{\v{c}}ek}, V., {et~al.} 2020,
  \planss, 184, 104871

\bibitem[{{Lyytinen} \& {Jenniskens}(2003)}]{lyy03}
{Lyytinen}, E. \& {Jenniskens}, P. 2003, \icarus, 162, 443

\bibitem[{{Miskotte}(2020)}]{mis20}
{Miskotte}, K. 2020, eMeteorNews, 5, 186

\bibitem[{{Pecina} \& {Ceplecha}(1983)}]{pec83}
{Pecina}, P. \& {Ceplecha}, Z. 1983, Bulletin of the Astronomical Institutes of
  Czechoslovakia, 34, 102

\bibitem[{{Roggemans} {et~al.}(2020){Roggemans}, {Howell}, \& {Gulon}}]{rog20}
{Roggemans}, P., {Howell}, J. A.~A., \& {Gulon}, T. 2020, eMeteorNews, 5, 13

\bibitem[{{Roggemans} {et~al.}(2018){Roggemans}, {Johannink}, \&
  {Biets}}]{rog18}
{Roggemans}, P., {Johannink}, C., \& {Biets}, J.-M. 2018, eMeteorNews, 3, 12

\bibitem[{{Shrben{\'y}} \& {Spurn{\'y}}(2019)}]{shr19}
{Shrben{\'y}}, L. \& {Spurn{\'y}}, P. 2019, \aap, 629, A137

\bibitem[{{Southworth} \& {Hawkins}(1963)}]{sou63}
{Southworth}, R.~B. \& {Hawkins}, G.~S. 1963, Smithsonian Contributions to
  Astrophysics, 7, 261

\bibitem[{{Spurn{\'y}} {et~al.}(2017){Spurn{\'y}}, {Borovi{\v{c}}ka}, {Mucke},
  \& {Svore{\v{n}}}}]{spu17}
{Spurn{\'y}}, P., {Borovi{\v{c}}ka}, J., {Mucke}, H., \& {Svore{\v{n}}}, J.
  2017, \aap, 605, A68

\bibitem[{{\v{S}tork} {et~al.}(1998){\v{S}tork}, {Borovi{\v{c}}ka},
  {Bo{\v{c}}ek}, \& {{\v{S}}olc}}]{sto98}
{\v{S}tork}, R., {Borovi{\v{c}}ka}, J., {Bo{\v{c}}ek}, J., \& {{\v{S}}olc}, M.
  1998, Meteoritics and Planetary Science Supplement, 33, A151

\bibitem[{{Voj{\'a}{\v{c}}ek} {et~al.}(2015){Voj{\'a}{\v{c}}ek},
  {Borovi{\v{c}}ka}, {Koten}, {Spurn{\'y}}, \& {{\v{S}}tork}}]{voj15}
{Voj{\'a}{\v{c}}ek}, V., {Borovi{\v{c}}ka}, J., {Koten}, P., {Spurn{\'y}}, P.,
  \& {{\v{S}}tork}, R. 2015, \aap, 580, A67

\bibitem[{{Voj{\'a}{\v{c}}ek} {et~al.}(2019){Voj{\'a}{\v{c}}ek},
  {Borovi{\v{c}}ka}, {Koten}, {Spurn{\'y}}, \& {{\v{S}}tork}}]{voj19}
{Voj{\'a}{\v{c}}ek}, V., {Borovi{\v{c}}ka}, J., {Koten}, P., {Spurn{\'y}}, P.,
  \& {{\v{S}}tork}, R. 2019, \aap, 621, A68

\bibitem[{{Znojil} \& {Hornoch}(1995)}]{zno95}
{Znojil}, V. \& {Hornoch}, K. 1995, WGN, Journal of the International Meteor
  Organization, 23, 205

\end{thebibliography}
\end{document}